# DEPLOYERS: An agent-based modeling tool for multi-country real-world data[1]

Martin Jaraiz and Ruth Pinacho, University of Valladolid, Spain

**Abstract**

We present recent progress in the design and development of DEPLOYERS, an agent-based macroeconomics modeling (ABM) framework, capable to deploy and simulate a full economic system (individual workers, goods and services firms, government, central and private banks, financial market, external sectors…) whose structure and activity analysis reproduce the desired calibration data, that can be, for example a Social Accounting Matrix (SAM) or a Supply-Use Table (SUT) or an Input-Output Table (IOT). Here we extend our previous work to a multi-country version and show an example using data from a 46-countries 64-sectors FIGARO Inter-Country IOT. The simulation of each country runs on a separate thread or CPU core to simulate the activity of one step (month, week, or day) and then interacts (updates imports, exports, transfers…) with that country's foreign partners, and proceeds to the next step. This interaction can be chosen to be aggregated (a single row and column IO account) or disaggregated (64 rows and columns) with each partner. A typical run simulates thousands of individuals and firms engaged in their monthly activity and then records the results, much like a survey of the country's economic system. This data can then be subjected to, for example, an Input-Output analysis to find out the sources of observed stylized effects as a function of time in the detailed and realistic modeling environment that can be easily implemented in an ABM framework.

## 1. Introduction

Policy making usually refers to decisions made at top levels, such as a government or a central bank. But the increasing availability of appropriate data and tools (software) is leading private companies and institutions to also use them as a valuable source of advice to help them make their decisions. This, in turn, can lead to better dynamic self-adjustment of the economic system and thus reduce frictions and alleviate the task of governments and central banks in helping to keep the system running smoothly.

---

[1] Presented at the 30th International Input-Output Association Conference, 2024, Santiago, Chile.

Ideally, in macroeconomics it would be better to operate at a more disaggregated level because, for example, some sectors may be more sensitive than others to a change in the interest rate or electricity tax. Again, from a modeling point of view, a macroeconomic system can be compared to a living being, i.e., a complex and constantly changing system composed of different interacting organs. Today, instrumentation plays a fundamental role in any hospital, although it would be useless without physicians. Similarly, it is desirable to improve the tools for macroeconomics to help economists understand why a complex system is malfunctioning, which sectors or subsectors are responsible and what can be done to bring it back on track without harming other sectors.

One reason for trying to introduce ABM into policy making is its potential to describe highly detailed economic structures and agent behaviors [1] and has been recommended as a tool worth to be explored [2]: "The atomistic, optimising agents underlying existing models do not capture behaviour during a crisis period. We need to deal better with heterogeneity across agents and the interaction among those heterogeneous agents. ... Agent-based modeling dispenses with the optimisation assumption and allows for more complex interactions between agents. Such approaches are worthy of our attention."

In fact, several agent-based macroeconomic modeling environments have already been developed (see [3] for a review). However, most of them have been used to conduct research in different specific areas of economic policy, and not as a general simulator of the real economy of a region or set of regions such as, for example, the European Union. Strikingly, none of the four main valid models for the euro area recently listed by [4] is agent-based. This work presents recent progress in the development of DEPLOYERS [5], an agent-based approach, following a Darwinian approach of survival of the fittest firms, which deploys and calibrates an economic system to reproduce, for example, a Social Accounting Matrix (SAM) or a Supply-Use Table (SUT) or an Input-Output Table (IOT) of a country, as we demonstrate using data from a FIGARO Inter-Country IOT of 46 countries with 64 sectors [6]. It should be noted that the ABM model used can, in principle, be arbitrary and that we have implemented a simple one only to test the calibration approach.

## 2. Description of the agent-based model used

As a basic example to describe the approach, we have chosen a SAM of Spain [7] with only six activity sectors and implemented a set of simple behavioral rules for the agents.

The implemented model (a modified version of the model described in [1]) simulates the activity of a population of households (active individuals) in a geographic region, initially without firms, with a government, a central bank (CB) and an external sector (rest of world, RoW). As the simulation progresses, some households will open firms of types (sectors) based on local demand, and some private banks. Also, a financial market will start as some firms reach a sufficiently large net worth. Unprofitable companies close and their owners look for work in their neighborhood. Prices evolve from interaction with neighbors following simple supply and demand rules.

Households initially have some monthly cash wages. They go out once every time step (month) on a random day to buy goods from the neighborhood, initially in the proportion given by the SAM (households consumption) but modulated with a logit probability as a function of prices. The consumption budget at step t, $C_{h,t}$, depends on the average income ($I_{h,t}$) and wealth ($W_{h,t}$) of each household, given by the following "buffer stock" rule:

$$C_{h,t} = I_{h,t} + \kappa \cdot (W_{h,t} - \phi \cdot I_{h,t}) \qquad (1)$$

where $\kappa$ is a sensitivity parameter and $\phi$ a buffer size. Households divide their surplus into bank deposits and risky assets (equity shares of individual firms) in the stock market.

Initially there are no firms, but households without a firm have a certain probability of opening a new one of the most frequently demanded types in the neighborhood. Firms can borrow from banks and have a random but fixed day of the month for their productive activity. They can sell their stock on any day to households or to other firms for intermediate consumption (IC). Each firm attempts to produce enough to replenish a level of inventories that is estimated based on recent demand. If the firm has liquidity needs to finance production (IC and taxes according to its SAM column ratios, plus labor and capital according to a Cobb-Douglas or Leontief model) it can apply for a bank loan and the subsequent production volume is conditioned to the outcome of the application. In addition, if a company meets the requirements to enter the stock market, it can also issue new shares. After taxes are paid, a fraction of the profits is distributed as dividends to the owner or shareholders, and the remainder is deposited in the payment account.

To account for all the economic activities, capital goods are also produced like consumption goods, regardless of their use: a new door can be used as gross fix capital formation (GFCF) for a new building or as a consumption good to replace another door in an old building. We use the coefficients in the GFCF column of the SAM to distribute a GFCF value into its goods and services components.

A basic Stock market model takes place in a clearing house which collects at the end of the month all the sell and buy orders (sorted from high to low price) and allocates them starting with the high price orders. Prices are readjusted following the supply and demand rule mentioned above.

Banks can be founded by households that meet the criteria defined by the Central Bank, such as a minimum initial net worth and a maximum number of banks. We have implemented the model described in [1], in which the bank's ability to extend credit is constrained by a capital adequacy requirement (CAR) and a reserve requirement ratio (RRR), but with a simple first-come, first-served response to loan applicants. The interest rate offered to a firm is an increasing function of credit risk, based on the firm's probability of default on the loan, which is estimated from its debt-to-equity ratio.

The government, according to the data provided by the SAM, collects taxes and redistributes them in the form of subsidies (unemployment) to households and public spending. In our simple model, the government uses the Central Bank to deposit or withdraw its surplus or deficit, respectively.

The external sector, like the government, is implemented as a simple input-output or consumer (SAM column) - producer (SAM row of IC for firms) account.

### 3. The self-deployment approach

Here we provide a summary of [5]. The deployment of the economic system is based on a simple Darwinian survival of the fittest approach. Initially there are no firms, only the active population, the government, the CB, and the external sector. To build up the production structure (number, fixed capital size, number of employees, and activity level of firms), it is initially established that the final consuming agents (households, government, and external sectors) repeatedly try to buy from neighbors their SAM value every month. During an initial deployment stage, individuals open firms corresponding to activity sectors of the SAM, and close unprofitable ones, until the production levels match the experimental values (output, unemployment, final and intermediate consumption). Producers also try to buy from the firms in their neighborhood the IC goods they need. The deployment stage ends when the economic system reaches the desired level of activity. Then follows a calibration stage, at constant demand levels, until the system reaches a steady state (constant average stock levels, unemployment...).

This configuration snapshot (the deployed, self-sustaining economic system) can be saved as an initial state to run different what-if tests. From this initial state on, the system is allowed to freely evolve according to the demand, supply and behavioral rules established by the model, without imposing the SAM's consumption quantities. For example, each household consumption budget follows eq. (1). For a fuller explanation, see Ref. 5.

### 4. Calibration with the inter-country FIGARO tables

This self-deployment approach has proven to achieve convergence to a stable economic system using larger IO tables with more than 30 sectors (Ref. 5). The DEPLOYERS implementation ca now be calibrated with intercountry databases like the 46-countries, 64-sectors FIGARO Input-Output Tables.

The simulation of each country runs on a separate thread (or CPU core) to simulate the activity of one step (month, week, or day) and then interacts (updates imports, exports, transfers…) with that country's foreign partners and proceeds to the next step. This interaction can be chosen to be aggregated (a single row and column) or disaggregated (64 rows and columns) for each partner. A typical run simulates thousands of individuals and firms engaged in their monthly activity and then records the results, very much like a survey of the country's economic system. This data can then be subjected to, for example, an Input-Output analysis to find out the sources of observed stylized effects as a function of time in the detailed and realistic modeling environment that can be easily implemented in an ABM framework.

As an example, a typical personal computer with 4 cores can run the simulation of Spain with France as disaggregated, and with Germany, the US, and the Rest of World (RoW) each as aggregated external sectors. The three countries are each simulated at a core. The CPU time is approximately 2 to 4 times the simulation of only Spain and the RoW, because of the 64 external sectors rows and columns from France (disaggregated). Memory requirements are not a limiting factor. Thus, hardware with 46 cores can run a World simulation with the 46 countries (plus the RoW, different but constant for each country), each interacting in detail with a few of its most active partners.

Figure 1 shows an example of deployment (up to step 120), calibration (up to step 240), and simulation of Spain (in this simulation, the other 35 countries were assembled with the RoW into a single exogenous sector) using the FIGARO intercountry input-output, industry by industry, 2018 table (64 sectors, 46 countries).

The households increase their consumption due to their increasing wealth. The gross output plot rises with demand up to step 360 but afterwards the output gap (real production minus its potential level) becomes increasingly negative because there is almost full employment (not shown). The economic system has reached its potential GDP output [8] because, in the simple model implemented here, there is constant technology. The subsequent exponential decline to zero can be explained by a positive feedback: to purchase their intermediate consumption (IC) goods, firms must compete with households and, since the economy is at full production, some firms fail to buy enough IC to produce and must close, and this aggravates the situation. Normally, the government or other agents take measures in advance to prevent a free fall of the economic system. For example, in August 2020, hundreds of thousands of Californians briefly lost power in rolling blackouts amid a heat wave, marking the first time outages were ordered in the state due to insufficient energy supplies in nearly 20 years [9].

For an active population of 3680 workers per country this simulation takes 2 CPU hours.

## 5. Conclusion

The methodology presented allows the use of agent-based models for real-world economies and, in principle, can be used to calibrate any ABM model using readily available inter country databases like FIGARO Input-Output Tables. This self-deployment and calibration methodology, based on a Darwinian survival of the fittest approach, opens the way to the use of ABM as a tool for policy makers dealing with complex data from real macroeconomic systems. It also enables the expected high performance of ABM models to address the complexities of current global macroeconomics, such as ecology, epidemiology, or social networks, as well as other areas of interest.

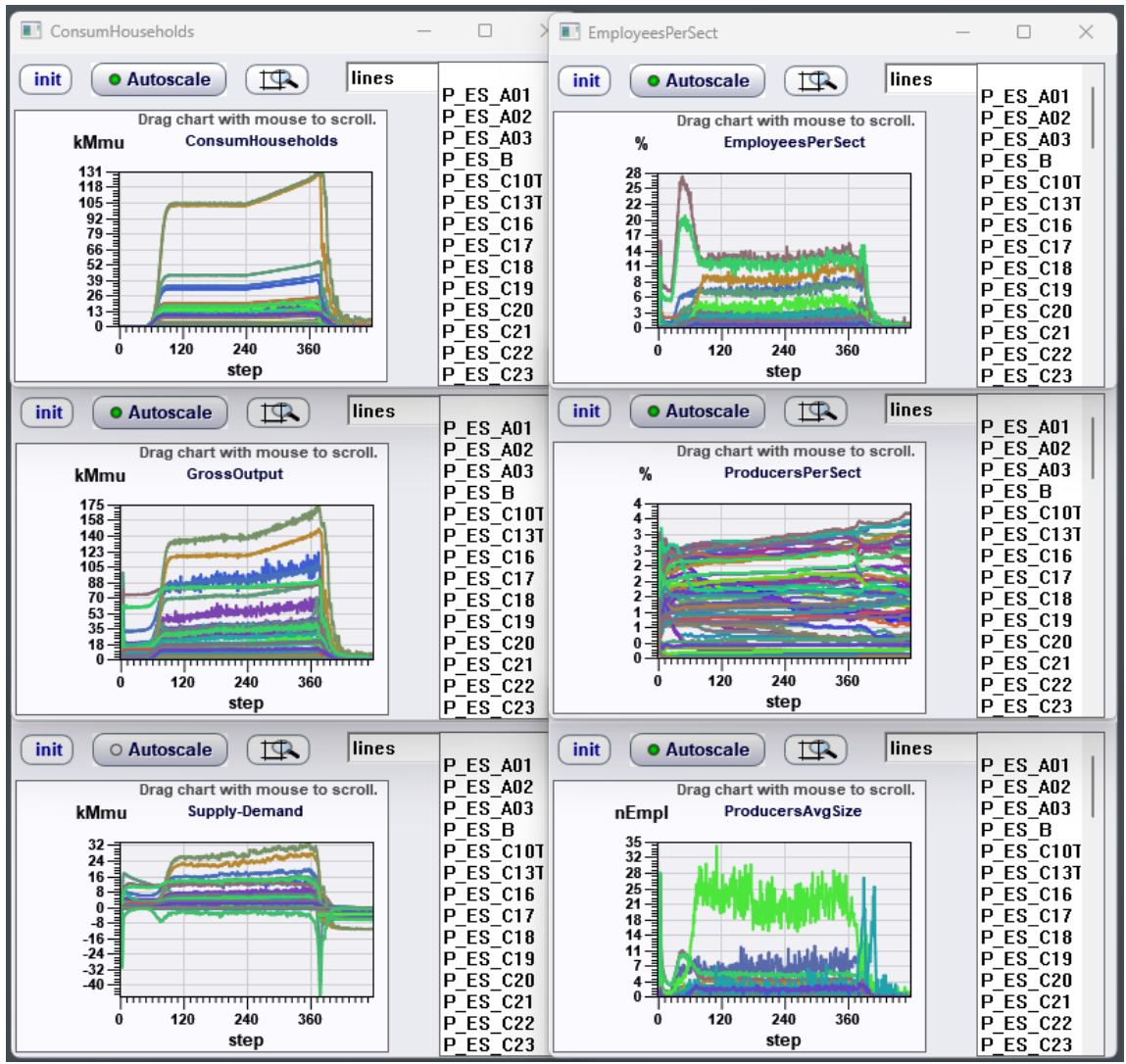

Figure 1. Example of deployment and calibration, up to month 240, of Spain (rest of World as a single external sector in this simulation) using the FIGARO intercountry input-output, industry by industry, 2018 table (64 sectors, 46 countries). The simulated active population is 3680, simulation time 2 CPU hours.

# DEPLOYERS

# An agent-based modeling tool
# for multi-country real-world data

Martin Jaraiz and Ruth Pinacho

*Dept. de Electronica, ETSI Telecomunicacion*
*Universidad de Valladolid, Spain*

**GLOBAL ECONOMY, ONE OF THE CURRENT MAJOR CHALLENGES**

- Policymakers complain about the lack of useful macroeconomic models and tools that help *at least to understand*, if not to predict, economic crises.

- Macroeconomic studies can be carried out through:
    1. Econometric analysis of data series.
    2. Input-Output matrix analysis.
    3. Simulation:
        1. Equation-based modeling, like Dynamic Stochastic General Equilibrium (DSGE).
        2. Simulation of the activity of agents (individuals, firms, banks,...) followed by data analysis, as if it were a Statistics Office survey (Agent-Based Modeling, ABM).

- Currently, macroeconomic simulation is dominated by DSGE.

- But DSGE can only reach a low level of disaggregation because the number of equations for current highly interconnected global economy becomes intractable.

# AGENT-BASED MODELING (1/2)

- Several agent-based macroeconomic modeling environments have already been developed.

- Agent-based simulators have been used to conduct research in specific areas of economic policy, but not as a general simulator of a real economy like, for example, the European Union. Examples of studies:

    - "Securitization and business cycle: an agent-based perspective"
    - "Macroprudential policies in an agent-based artificial economy"
    - "Macroeconomic implications of mortgage loans requirements: An agent-based approach".

- But none of the four main valid models for the euro area, recently listed by Blanchard, is agent-based (*Macroeconomics, a European perspective*, 2021).

Why?

**AGENT-BASED MODELING (2/2)**

- The main hindrance to the use of ABM is the difficulty to set up an *initial economic system* (labor, number and size of firms, a central bank, the government, external sectors...) that reproduces the structure and activity of a real system like a country.

- Despite efforts, that include Artificial Intelligence techniques, the estimation or calibration of model parameters remains a challenge and a roadblock to bring ABM to real economic systems.

The DEPLOYERS program: a Darwinian approach to deploy real-world agent-based economic systems

- The ABM approach presented here can read a SAM, IOT or SUT and deploy from scratch an economic system (labor, sectors of activity operating as firms, a central bank, the government, external sectors...) whose structure and activity yield a SAM with values in close agreement with those of the actual SAM snapshot.

- What is important here is not the specific agent-based model used in this example (which can be any other) but the ability to deploy an economic structure that generates a SAM with the desired values.

A real economic system can be described by a Social Accounting Matrix

like this simple one (**six sectors**) for Spain in 2008

Firms are created <u>following the pattern </u>of each sector: when a firm from the

AgroPesc sector sells 48021 product units pays 4259 m.u. as wages.

| | A | B | C | D | E | F | G | H |
|---|---|---|---|---|---|---|---|---|
| 1 | SAM_table { MCAESP08 | | | | | | | |
| 2 | SPAIN | Year: 2008 | | Population: 40000000 | | | Active: 20000000 | InitUnemp: |
| 3 | | P01_AgroPesc | P02_EnerPetro | P03_Indust | P04_Construc | P05_ServVenta | N06_ServNoVenta | F07_GFCF |
| 4 | P01_AgroPesc | 1701 | 1 | 24972 | 24 | 2877 | 302 | 811 |
| 5 | P02_EnerPetro | 1119 | 41384 | 19205 | 2678 | 21061 | 6438 | 292 |
| 6 | P03_Indust | 8616 | 5037 | 209653 | 64805 | 76816 | 21507 | 65355 |
| 7 | P04_Construc | 190 | 793 | 1909 | 108372 | 25997 | 3654 | 176136 |
| 8 | P05_ServVenta | 3682 | 10762 | 94805 | 31726 | 244738 | 49376 | 54460 |
| 9 | N06_ServNoVenta | 0 | 0 | 0 | 0 | 0 | 0 | 0 |
| 10 | F07_GFCF | 0 | 0 | 0 | 0 | 0 | 0 | 0 |
| 11 | X08_SectExt | 8742 | 35259 | 238669 | 804 | 60405 | 1085 | 0 |
| 12 | L09_CompEmployees | 4259 | 4220 | 62958 | 50681 | 200109 | 88364 | 0 |
| 13 | K10_GrossOpSurplus | 19802 | 17740 | 48917 | 46113 | 321169 | 14029 | 0 |
| 14 | T11_SSoc | 624 | 1543 | 19007 | 15518 | 53763 | 26223 | 0 |
| 15 | T12_TaxProduction | -244 | 120 | -137 | 204 | 1118 | 53 | 0 |
| 16 | T13_TaxProducts | -471 | 307 | -1062 | 1555 | 12274 | 4100 | 21578 |
| 17 | T14_IRPF | 0 | 0 | 0 | 0 | 0 | 0 | 0 |
| 18 | G15_Government | 0 | 0 | 0 | 0 | 0 | 0 | 0 |
| 19 | H16_Households | 0 | 0 | 0 | 0 | 0 | 0 | 0 |
| 20 | colSUM | 48021 | 117166 | 718896 | 322480 | 1020327 | 215131 | 318632 |

Wages

Intermediate
consumption

Value
added

Dividends
+ GFCF

Gross
output

We obtain the
included GFCF
component

Firms act <u>in response to the demand</u> of the Final Consumers and of other firms (for their Intermediate consumption).

Final consumers

| X08_SectExt | L09_CompEmplo | K10_GrossOpSur | T11_SSoc | T12_TaxProduction | T13_TaxProducts | T14_IRPF | G15_Government | H16_Households |
|---|---|---|---|---|---|---|---|---|
| 7834 | 0 | 0 | 0 | 0 | 0 | 0 | 0 | 9499 |
| 6740 | 0 | 0 | 0 | 0 | 0 | 0 | 0 | 18249 |
| 144645 | 0 | 0 | 0 | 0 | 0 | 0 | 7620 | 114842 |
| 144 | 0 | 0 | 0 | 0 | 0 | 0 | 0 | 5285 |
| 70689 | 0 | 0 | 0 | 0 | 0 | 0 | 21238 | 438851 |
| 0 | 0 | 0 | 0 | 0 | 0 | 0 | 211972 | 3159 |
| 103916 | 0 | 0 | 0 | 0 | 0 | 0 | 12014 | 202702 |
| 0 | 0 | 0 | 0 | 0 | 0 | 0 | 0 | 0 |
| 0 | 0 | 0 | 0 | 0 | 0 | 0 | 0 | 0 |
| 0 | 0 | 0 | 0 | 0 | 0 | 0 | 0 | 0 |
| 0 | 0 | 0 | 0 | 0 | 0 | 0 | 0 | 20074 |
| 0 | 0 | 0 | 0 | 0 | 0 | 0 | 0 | 0 |
| -92 | 0 | 0 | 0 | 0 | 0 | 0 | 401 | 53758 |
| 0 | 0 | 0 | 0 | 0 | 0 | 0 | 0 | 117483 |
| 0 | 0 | 0 | 136752 | 1114 | 92348 | 117483 | 0 | 0 |
| 11088 | 410591 | 467771 | 0 | 0 | 0 | 0 | 94452 | 0 |
| 344964 | 410591 | 467771 | 136752 | 1114 | 92348 | 117483 | 347697 | 983902 |

# Agent-based modeling of real-world economic systems

 Approach followed for the optimization task of finding a suitable economic structure:

The difficulty lays on <u>finding a stable, self-sustaining</u> economic system: number and size of each type of firm, household wealth and demand distribution, ...

*Problem to be solved*: calibration of parameters

<u>*Proposed solution*</u>: "let the agents do the job" (as they do in real life), no maths

*Approach*: survival of the firms best fitted to their supply-demand environment

# The deployment approach: An overview (1/2)

- To build the economic system, the approach simulates the activity of a population of **households** (active population) in a geographic region, initially with a **government**, a **central bank** (CB) and an **external sector**.

- The simulation starts without any firm.

- Households begin to try to buy goods and open firms, of types chosen based on <u>local demand</u>.

- Survival of the fittest: Firms that become unprofitable are closed.

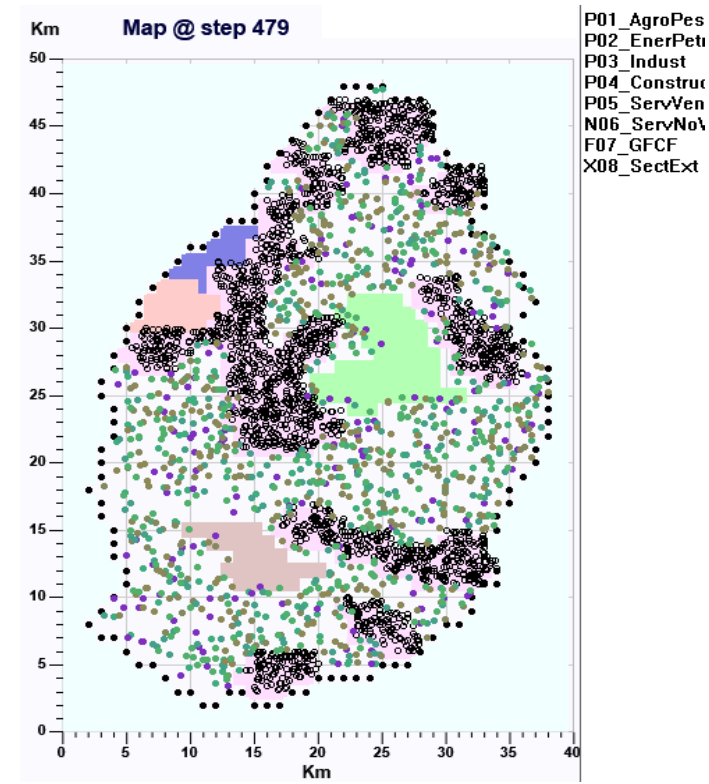

# The deployment approach: An overview (2/2)

- A **financial market** will start as some firms reach a sufficiently large net worth (or meet any other requirement).

- **Prices** evolve from interaction with neighbors following simple **supply-demand rules**: if the buyer's price is greater than or equal to the seller's price, the transaction takes place, and the buyer's price decreases and the seller's price increases by a small factor. Otherwise, there is no transaction and prices change in the opposite direction.

- For the simple model and data (a **six sectors SAM**) used in the example, **2000** (active) **individuals** are sufficient to obtain statistically acceptable results and it takes about **2 minutes** on standard CPUs. Since agents can have a limited number of interactions per month, the **CPU time increases almost linearly with the simulated population**.

# Description of the model chosen for the deployment example (1/6)

(see Dawid et al. 2019)

- Households initially have a **few monthly cash wages**. They go out once every time step (month) on a random day **to buy goods** from the neighborhood, *initially in the relative proportion given by the SAM* (households consumption) but modulated with a <u>logit</u> probability as a function of prices.

- The consumption budget at step t, $C_{h,t}$, depends on the average income ($I_{h,t}$) and wealth ($W_{h,t}$) of each household (h), given by the following "buffer stock" rule:

$$C_{h,t} = I_{h,t} + K \cdot (W_{h,t} - \varphi \cdot I_{h,t})$$

- **Initially** there are **no firms**, but households without a firm have a certain probability of opening a new one of the most frequently demanded types in the neighborhood.

- Unlike households, **firms can borrow from banks**. They sell to households and to other firms for their **intermediate consumption (IC)**. Each firm attempts to produce enough to replenish a level of **inventories** that is **estimated based on average recent demand**.

- If the firm has liquidity needs to finance production, it can apply for a bank loan. Subsequent volume production is conditioned to the outcome of the application. In addition, **if a company meets the requirements to enter the stock market, it can issue new shares**.

- **After taxes are paid**, a fraction of the profits is distributed as **dividends** to the owner or shareholders, and the remainder is deposited in a payment account.

- A **simplified financial market** takes place in a clearing house that collects at the end of the month all sell and buy orders (ordered from high to low price) and allocates them starting from their high price end. **Prices are adjusted following the supply - demand rule mentioned above**.

# Description of the model chosen for the deployment example (4/6)

- **Banks** can be founded by households that meet the criteria defined by the Central Bank. We have followed the model described by Dawid et al. [4], in which **the bank's ability to extend credit is constrained by a capital adequacy requirement (CAR) and reserve requirement ratio (RRR)**. **The interest rate offered to a firm is an increasing function of credit risk**, following the internal risk-based (IRB) approach of the Basel Accords, based on the firm's probability of default on the loan, which is estimated from its debt-to-equity ratio.

- The **Government**, according to the data provided by the SAM, **collects taxes and redistributes them** in the form of subsidies (unemployment) to households and public spending. In our simple model, the government uses the Central Bank to deposit or withdraw its surplus or deficit, respectively.

- **"External sectors"** are implemented as special Import/Export **domestic firms**.

- These interface firms can **interact** (buy/sell) not only with **domestic** firms but **also** with external-sector interface firms from **other countries**.

- In the current implementation, they can interact with domestic companies and households **daily** and with other countries on a **monthly** basis.

## Building a SAM from FIGARO

The Import/Export interaction between two countries can chosen to be:

**Disaggregated (more CPU time)**

one sector <u>per domestic sector</u> in each country
(64 columns, 64 rows per external country)

**Aggregated**

one sector in each country
(1 column, 1 row per external country)

- During an initial **"deployment stage"**, individuals **open new firms** of the SAM activity sectors, and **close unprofitable ones**, until the activity output level reaches the experimental **final consumers** values.

Household consumption by sector

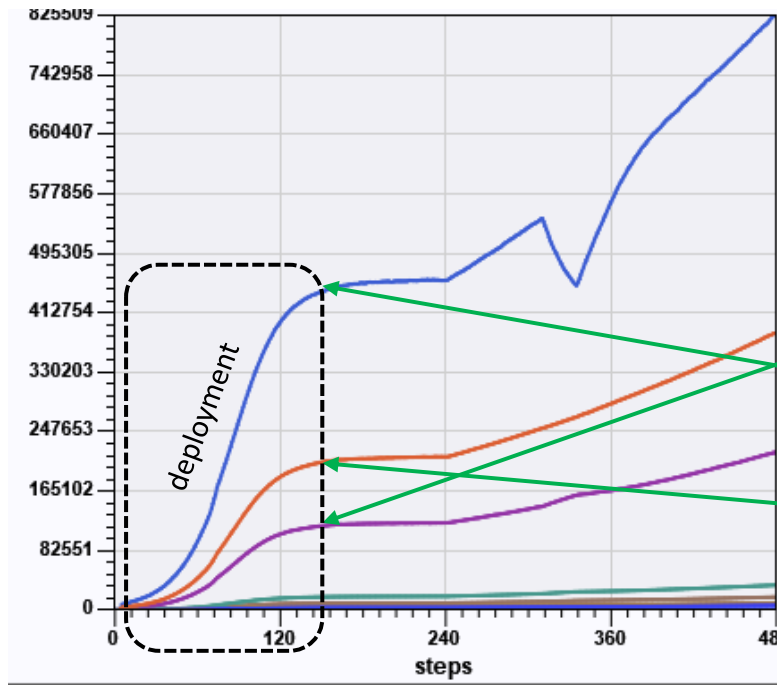

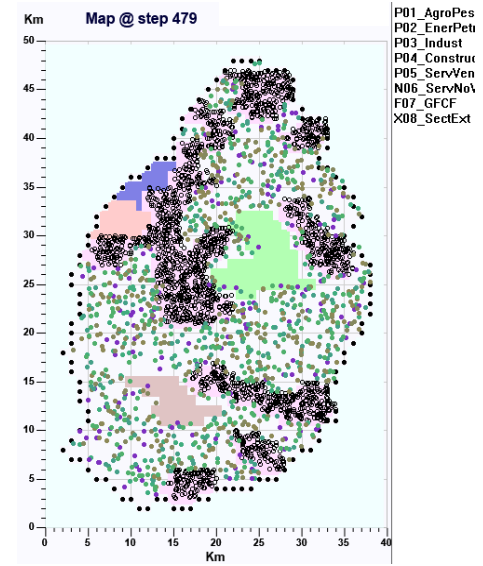

# THE SELF-DEPLOYMENT APPROACH (2/3)

- Upon deployment, the system undergoes a **"calibration stage"** until it reaches a steady state (like constant average stock of firms, not shown).

- During this stage, the program adjusts parameters like K in the household consumption budget.

- We refer to this deployment + calibration range the **"assisted production"** stage.

Household consumption by sector

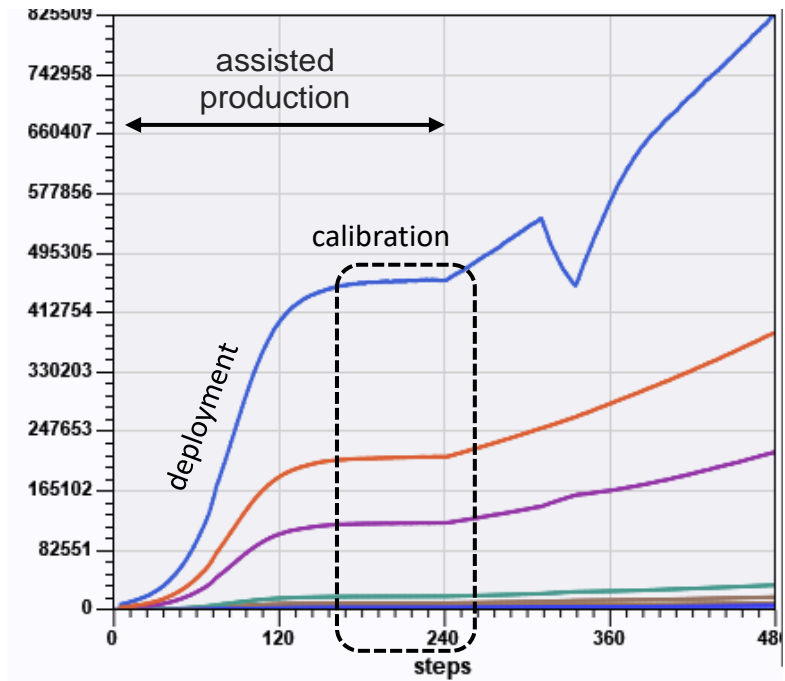

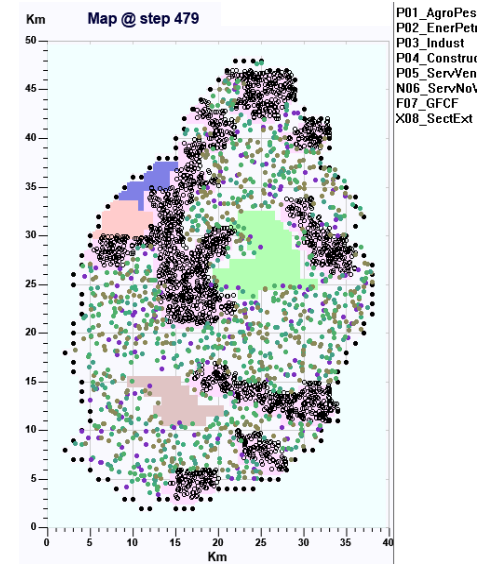

# THE SELF-DEPLOYMENT APPROACH (3/3)

- At the end of this "assisted production" stage (deployment + calibration) we have **the initial state** of an economic system that reproduces the values of the real system SAM.

- This state can be saved and **used to see the response of the system to a series of *what-if* tests**.

Household consumption by sector

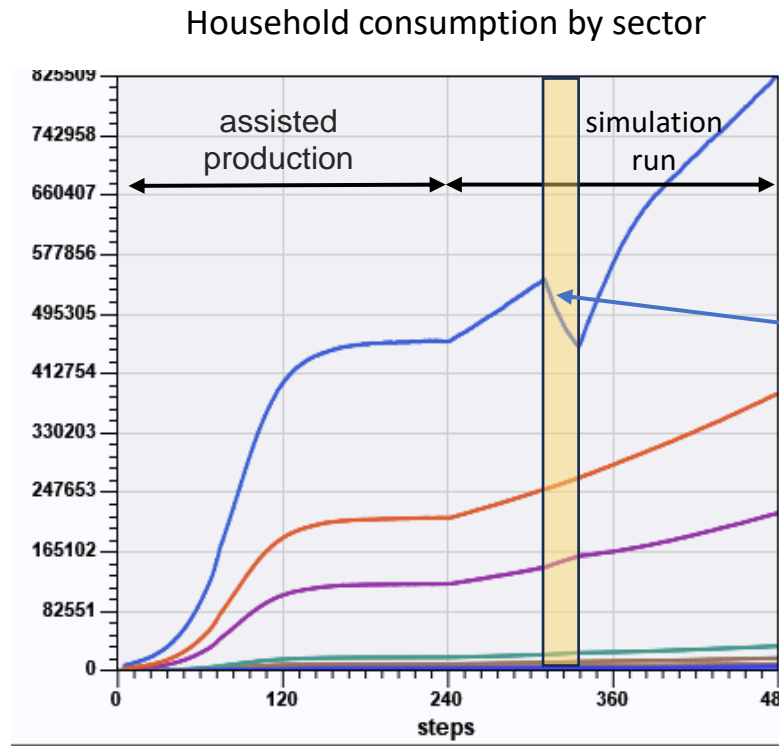

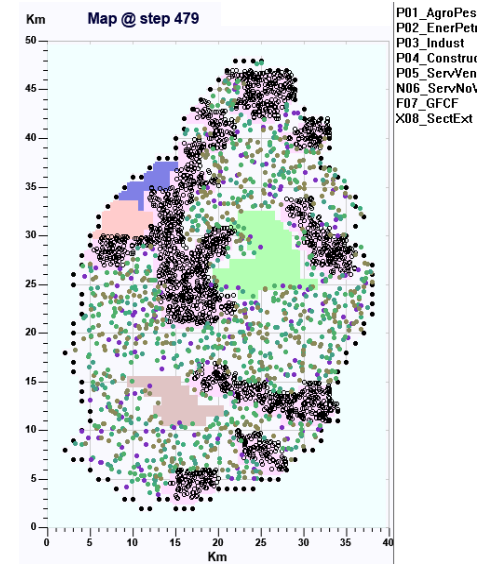

# EXAMPLES

**6-sectors** SAM, 2000 indivs

| | A | B | C | D | E | F | G | H | I | J | K | L | M | N | O | P | Q | R | S |
|---|---|---|---|---|---|---|---|---|---|---|---|---|---|---|---|---|---|---|---|
| 1 | SAM_table ( MCAESP08 | | | | | | | | | | | | | | | | | | |
| 2 | SPAIN | Year: 2008 | | Population: 40000000 | | Active: 20000000 | | InitUnemp: 12 | | Nproducers: 8 | | Naccounts: 16 | | | Units: 1000000 | euros | | | |
| 3 | | P01_AgroPesc | P02_EnerPetro | P03_Indust | P04_Construc | P05_ServVenta | N06_ServNoVenta | F07_GFCF | X08_SectExt | L09_CompEmplo | K10_GrossOpSur | T11_SSoc | T12_TaxProduction | T13_TaxProducts | T14_IRPF | G15_Government | H16_Households | rowSUM | |
| 4 | P01_AgroPesc | 1701 | 1 | 24972 | 24 | 2877 | 302 | 811 | 7834 | 0 | 0 | 0 | 0 | 0 | 0 | 0 | 9499 | 48021 | P01_AgroPesc |
| 5 | P02_EnerPetro | 1119 | 41384 | 19205 | 2678 | 21061 | 6438 | 292 | 6740 | 0 | 0 | 0 | 0 | 0 | 0 | 0 | 18249 | 117166 | P02_EnerPetro |
| 6 | P03_Indust | 8616 | 5037 | 209653 | 64805 | 76816 | 21507 | 65355 | 144645 | 0 | 0 | 0 | 0 | 0 | 0 | 7620 | 114842 | 718896 | P03_Indust |
| 7 | P04_Construc | 190 | 793 | 1909 | 108372 | 25997 | 3654 | 176136 | 144 | 0 | 0 | 0 | 0 | 0 | 0 | 0 | 5285 | 322480 | P04_Construc |
| 8 | P05_ServVenta | 3682 | 10762 | 94805 | 31726 | 244738 | 49376 | 54460 | 70689 | 0 | 0 | 0 | 0 | 0 | 0 | 21238 | 438851 | 1020327 | P05_ServVenta |
| 9 | N06_ServNoVenta | 0 | 0 | 0 | 0 | 0 | 0 | 0 | 0 | 0 | 0 | 0 | 0 | 0 | 0 | 211972 | 3159 | 215131 | N06_ServNoVenta |
| 10 | F07_GFCF | 0 | 0 | 0 | 0 | 0 | 0 | 0 | 103916 | 0 | 0 | 0 | 0 | 0 | 0 | 12014 | 202702 | 318632 | F07_GFCF |
| 11 | X08_SectExt | 8742 | 35259 | 238669 | 804 | 60405 | 1085 | 0 | 0 | 0 | 0 | 0 | 0 | 0 | 0 | 0 | 0 | 344964 | X08_SectExt |
| 12 | L09_CompEmployees | 4259 | 4220 | 62958 | 50681 | 200109 | 88364 | 0 | 0 | 0 | 0 | 0 | 0 | 0 | 0 | 0 | 0 | 410591 | L09_CompEmployees |
| 13 | K10_GrossOpSurplus | 19803 | 17740 | 48917 | 46113 | 321169 | 14029 | 0 | 0 | 0 | 0 | 0 | 0 | 0 | 0 | 0 | 0 | 467771 | K10_GrossOpSurplus |
| 14 | T11_SSoc | 624 | 1543 | 19007 | 15518 | 53763 | 26223 | 0 | 0 | 0 | 0 | 0 | 0 | 0 | 0 | 0 | 20074 | 136752 | T11_SSoc |
| 15 | T12_TaxProduction | -244 | 120 | -137 | 204 | 1118 | 53 | 0 | 0 | 0 | 0 | 0 | 0 | 0 | 0 | 0 | 0 | 1114 | T12_TaxProduction |
| 16 | T13_TaxProducts | -471 | 307 | -1062 | 1555 | 12274 | 4100 | 21578 | -92 | 0 | 0 | 0 | 0 | 0 | 0 | 401 | 53758 | 92348 | T13_TaxProducts |
| 17 | T14_IRPF | 0 | 0 | 0 | 0 | 0 | 0 | 0 | 0 | 0 | 0 | 0 | 0 | 0 | 0 | 0 | 117483 | 117483 | T14_IRPF |
| 18 | G15_Government | 0 | 0 | 0 | 0 | 0 | 0 | 0 | 0 | 0 | 0 | 136752 | 1114 | 92348 | 117483 | 0 | 0 | 347697 | G15_Government |
| 19 | H16_Households | 0 | 0 | 0 | 0 | 0 | 0 | 0 | 11088 | 410591 | 467771 | 0 | 0 | 0 | 0 | 94452 | 0 | 983902 | H16_Households |
| 20 | colSUM | 48021 | 117166 | 718896 | 322480 | 1020327 | 215131 | 318632 | 344964 | 410591 | 467771 | 136752 | 1114 | 92348 | 117483 | 347697 | 983902 | | |

100 * simulation / real values

SAM calculated after absoluteStep 360: ------------------------------------

| | P01_AgroPes | P02_EnerPet | P03_Indust | P04_Constru | P05_ServVen | N06_ServNoV | F07_GFCF | X08_SectExt | | G15_Governm | H16_Househo | |
|---|---|---|---|---|---|---|---|---|---|---|---|---|
| P01_AgroPesc | 100 | 55 | 102 | 117 | 105 | 91 | 108 | 100 | | | 105 | P01_AgroPesc |
| P02_EnerPetro | 100 | 97 | 102 | 106 | 105 | 91 | 102 | 100 | | | 105 | P02_EnerPetro |
| P03_Indust | 100 | 97 | 102 | 106 | 105 | 91 | 110 | 100 | | 100 | 105 | P03_Indust |
| P04_Construc | 97 | 97 | 102 | 106 | 105 | 91 | 111 | 100 | | | 104 | P04_Construc |
| P05_ServVenta | 100 | 97 | 102 | 106 | 105 | 91 | 110 | 100 | | 100 | 105 | P05_ServVenta |
| N06_ServNoVenta | | | | | | | | | | 100 | 106 | N06_ServNoVenta |
| F07_GFCF | | | | | | | | 100 | | 100 | 105 | F07_GFCF |
| X08_SectExt | 100 | 97 | 102 | 107 | 105 | 91 | | | | | | X08_SectExt |
| L09_CompEmployees | 92 | 83 | 95 | 99 | 96 | 89 | | | | | | L09_CompEmployees |
| K10_GrossOpSurplu | 100 | 92 | 104 | 109 | 105 | 98 | | | | | | K10_GrossOpSurplu |
| T11_SSoc | 100 | 92 | 104 | 109 | 105 | 98 | | | | 100 | | T11_SSoc |
| T12_TaxProduction | 100 | 92 | 104 | 109 | 105 | 98 | | | | | | T12_TaxProduction |
| T13_TaxProducts | 100 | 92 | 104 | 109 | 105 | 98 | | 100 | | 100 | | T13_TaxProducts |
| T14_IRPF | | | | | | | | | | 100 | | T14_IRPF |
| G15_Government | | | | | | | | | | | | G15_Government |
| H16_Households | | | | | | | | | | | | H16_Households |
| | P01_AgroPes | P02_EnerPet | P03_Indust | P04_Constru | P05_ServVen | N06_ServNoV | F07_GFCF | X08_SectExt | | G15_Governm | H16_Househo | |

# DEPLOYMENT EXAMPLES (1/4)

## Individuals Wealth distribution

### Simulation

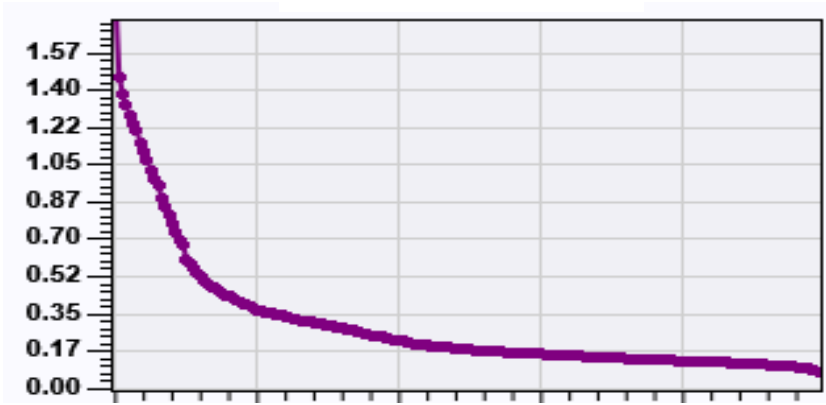

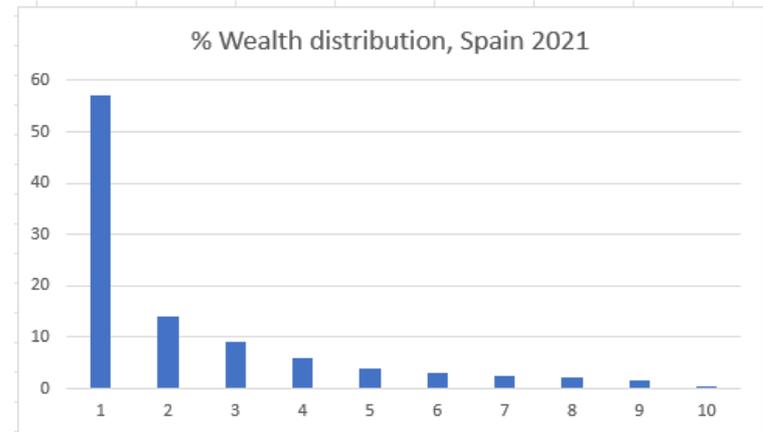

For the simple model and data used in the example (a **six sectors SAM**):

- **2000** (active) **individuals** are sufficient to obtain statistically acceptable results.

- It takes about **2 minutes** on standard CPUs. Since agents can have a limited number of interactions per month, the **CPU time increases almost linearly with the simulated population**.

# DEPLOYMENT EXAMPLES (1/4)

**In agreement with data**

**6-sectors** SAM, 2000 indivs [deployment & calibration upto 360]

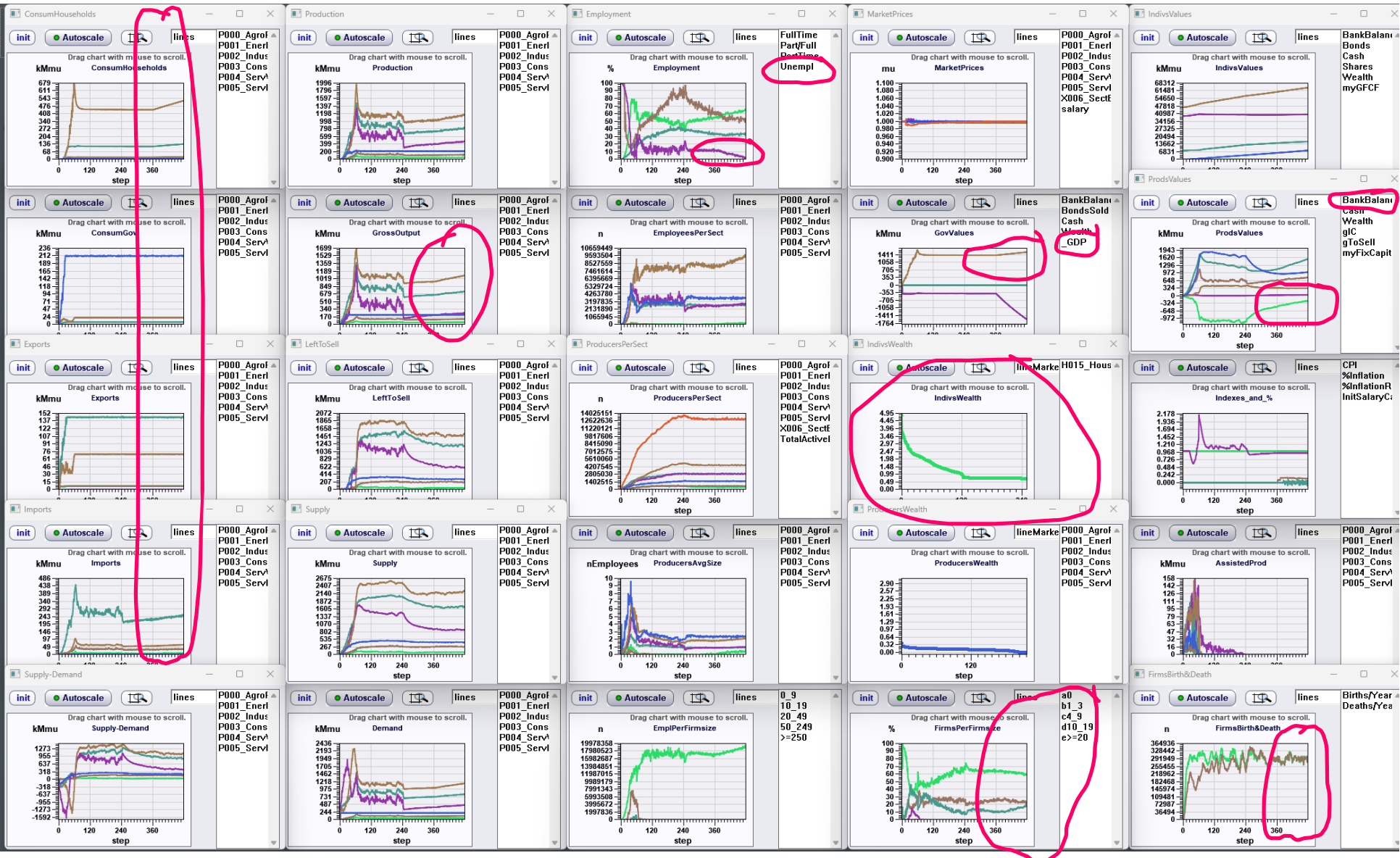

**In agreement with data**

Same 6-sectors SAM, but now disaggregated into 31-sectors, 2000 indivs

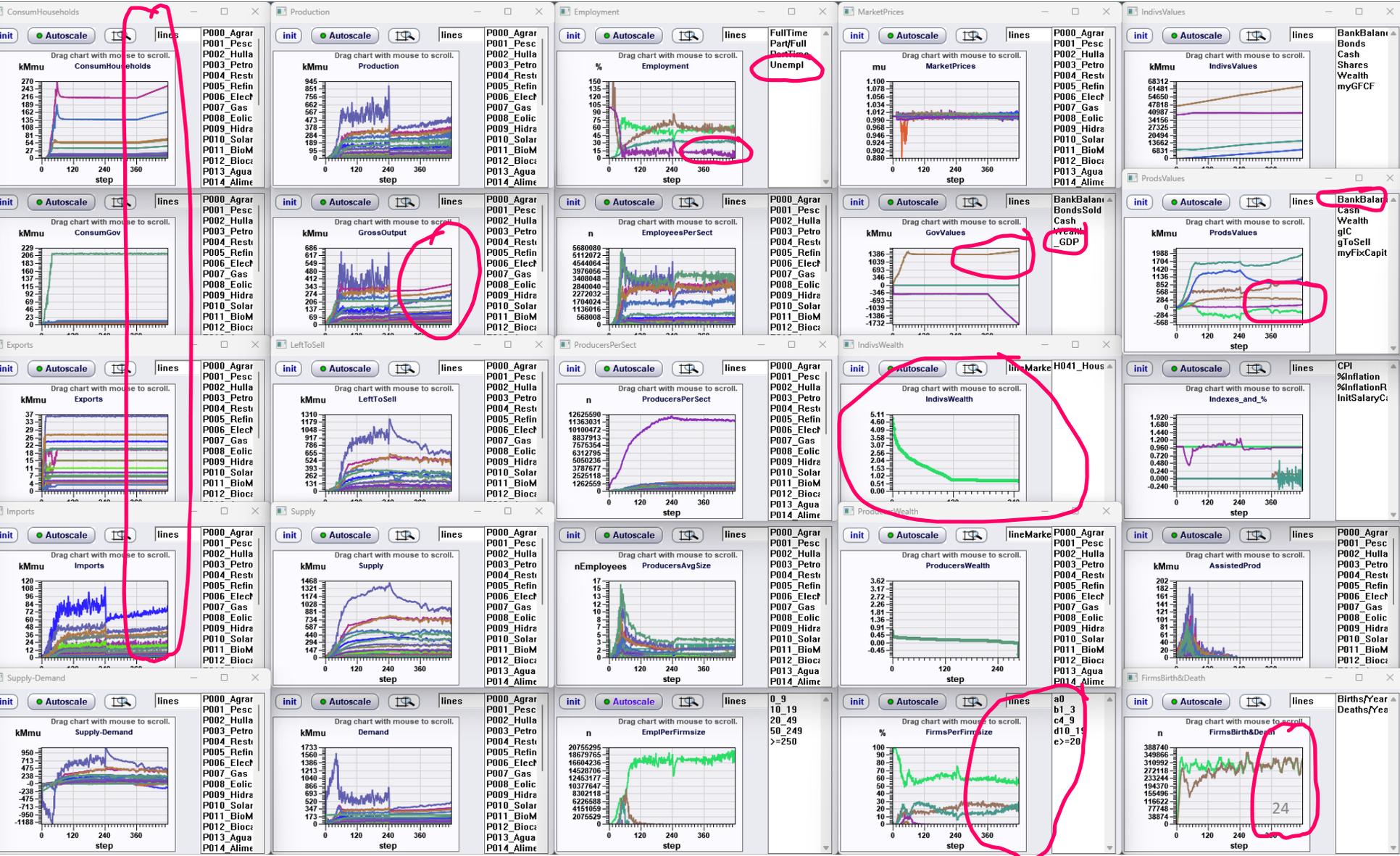

Deployment and calibration [upto 240] with a **Supply-Use Table**

P001_CropAnimProduction;;;;;;;;;;;;;;;;;;;;;;;;;;;;;;;34976;15199;671;11;;;;;;;519;257;375;59;129;246;;;25;84;
P002_Forestry;;;;;;;;;;;;;;;;;;;;;;;;;;;;;;;10;0;0;1837;;;;;0;;;;;;;;3;2;
P003_Fishing;;;;;;;;;;;;;;;;;;;;;;;;;;;;;;;2744;;;;;;;;;32;6;
P004_Mining;;;;;;;;;;;;;;;;;;;;;;;;;;;;;64;1103;18;936;2857;;;;;;;;21;11;
P005_MeatProcessing;;;;;;;;;;;;;;;;;;;;;;;;;;;;25927;58;;900;307;;;;49;62;
P006_DairyMnf;;;;;;;;;;;;;;;;;;;;;;;;;;;;;;;7841;;;473;;447;;;24;12;
P007_OtherFoodMn
P008_BeverageMnf
P009_TobaccoMnf;
P010_TextileMnf;
P011_WearingMnf;
P012_LeatherMnf;
P013_WoodMnf;;;
P014_PaperMnf;;
P015_Printing;;;
P016_PetrolMnf;;
P017_ChemMnf;;;;
P018_PharmaMnf;;
P019_PlasticMnf;
P020_NonmetalMn
F021_GFCF;;;;;;;
X022_ExtSectEU;;
X023_ExtSectNonE
L024_CompEmploye
K025_GrossOpSurp
T026_SSocEmploye
T027_TaxProducti
G028_Government;
H029_Households;
C030_Agric;**1621**;
C031_LiveAnim;**20**
C032_AgricSrv;**59**
C033_Forest;**3**;**73**
C034_Fish;;;**8**;**0**
C035_Coal;;;;**2**;;
C036_Crudepetrol
C037_NaturalGas;
C038_MetalOres;;
C039_OtherMining
C040_PreservMeat
C041_Dairy;**120**;;
C042_VegetOils;;
C043_PreparedFee

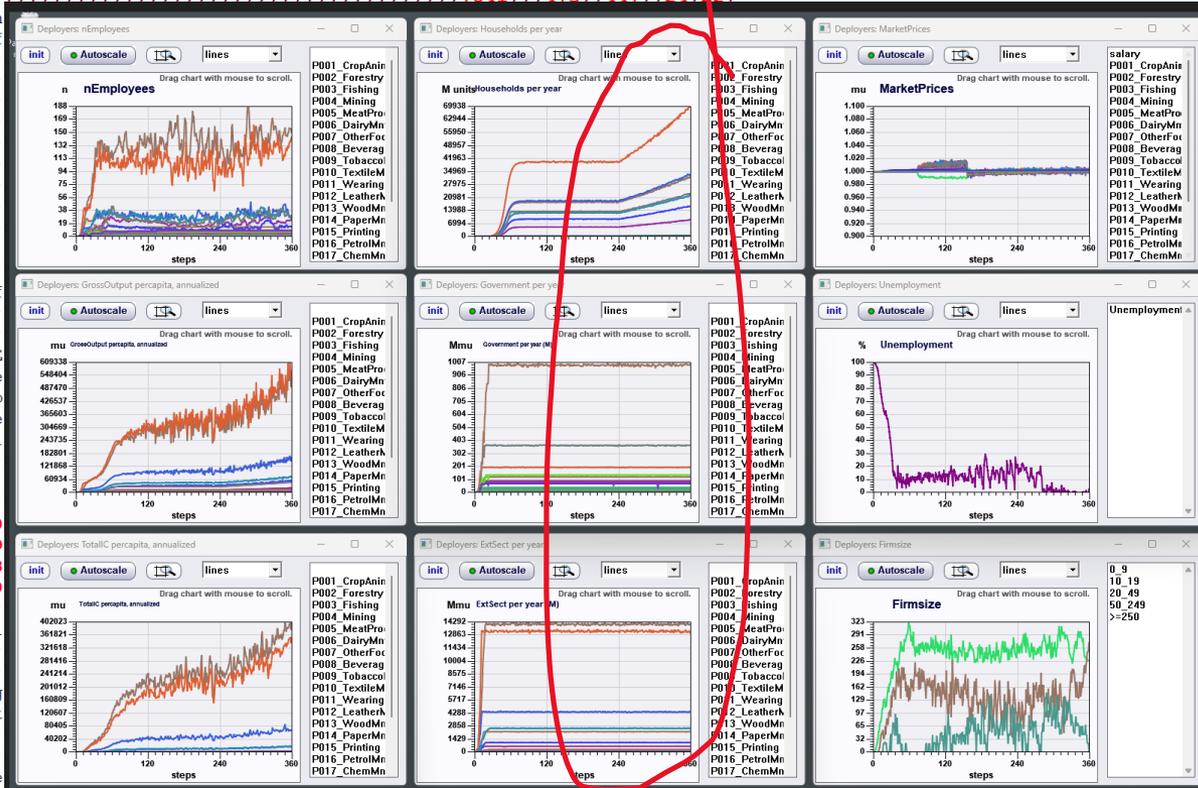

C044_OtherFood;**19**;;**55**;**35**;**180**;**133**;**7049**;**2267**;**0**;**1**;;;**1**;**38**;;;**32**;**5**;;;;**10324**;**4108**;;;;;**34393**;;;;;;;;;;;;;;;;
C045_Alcoholic;;;;**3**;**2**;**243**;**1306**;;;;**0**;;;;**18**;;;;**2346**;**1741**;;;;;**8098**;;;;;;;;;;;;;;;;;;;;;;
C046_NonAlcoholic;;;**1**;;;**1**;**324**;**258**;;;;;;;;;;**5**;;;;**236**;**86**;;;;**4833**;;;;;;;;;;;;;;;
C047_Tobacco;;;;;;;;;**7**;;;;;;;;;;;**240**;**54**;;;;;**13128**;;;;;;;;;;;;;;;;;;
C048_RepairServ;**313**;**14**;**158**;**110**;**122**;**40**;**299**;**144**;**2**;**34**;**9**;**11**;**161**;**220**;**31**;**133**;**308**;**103**;**132**;**240**;**10015**;**263**;**208**;;;;;;**322**;;;;;;;;;;;;;;;;;;;;;
C049_ScientificSrv;;;;;;;;;;;;;;;;;;;;**15579**;**641**;**514**;;;;;**2307**;;;;;;;;;;;;;;;;;;;;;;

## DEPLOYMENT EXAMPLES (4/4)

Using FIGARO intercountry tables

- Calibration to the FIGARO intercountry input-output, industry by industry, 2018 table (64 sectors, 46 countries)
- Example of deployment and calibration, up to month 360, of Spain and Portugal (aggregated External Sectors in this simulation)
- The simulated active population is only 500 individuals per country: enough to achieve a stable economic system and accurate GDP and some Final Consumptions (noisy imports for sim. size).
- 15 minutes CPU time.

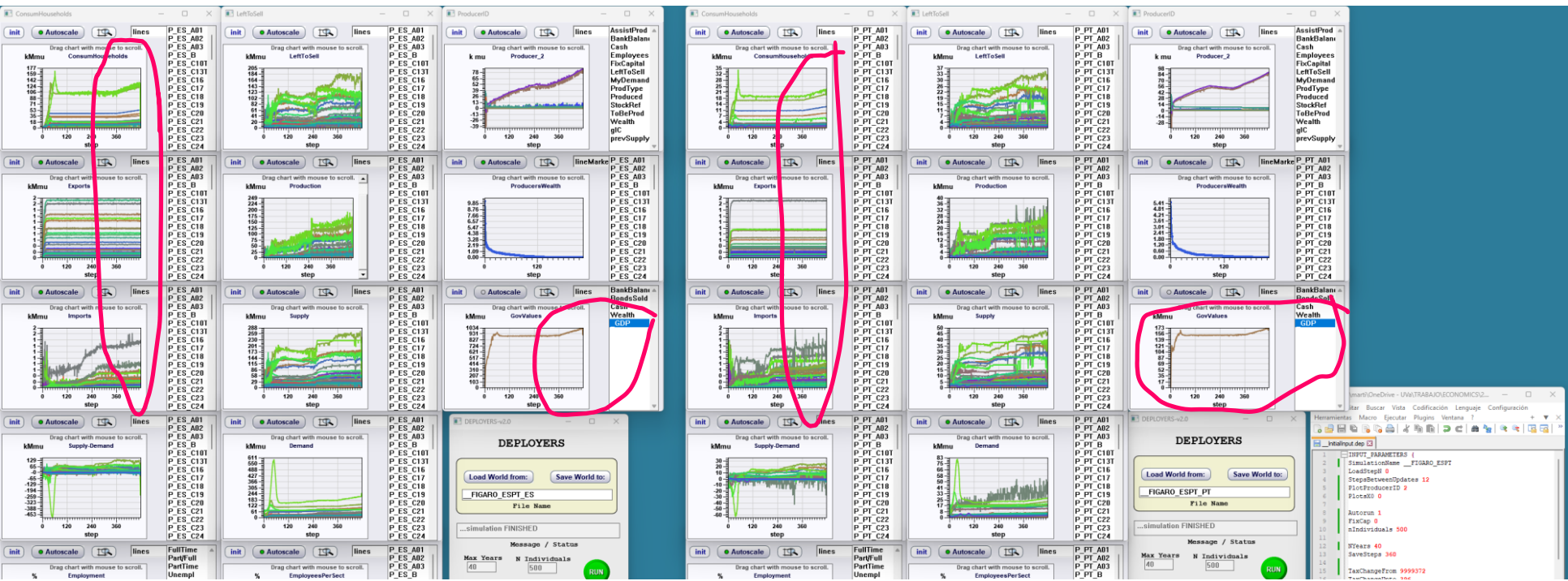

Besides FIGARO, **other databases** could, in principle, be used to calibrate **any ABM model** using this approach, <u>without user assistance</u>.

1. For example, the **OECD**, Inter-Country Input-Output (ICIO), industry by industry, tables (45 industries, 77 countries).[<u>VA needs to be split as VA = L + K</u>]

2.  Another example of <u>readily usable</u> database is the Supply-Use Tables

from the **U.S. Bureau of Economic Analysis**:

（

- <u>Because of its Agent-based</u> foundations, other layers from extended databases, like **ecology, social networks or epidemiology**, that can <u>interact with economics</u>, could be added in the future.

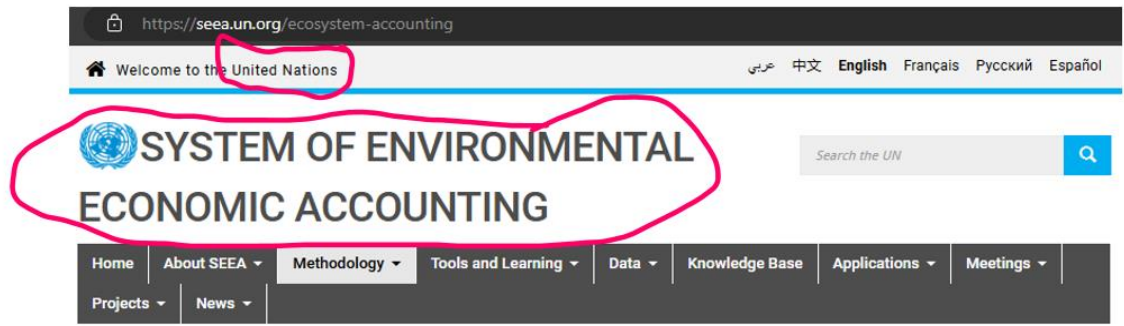

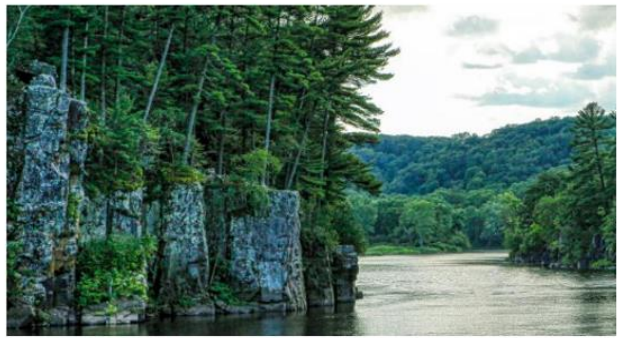

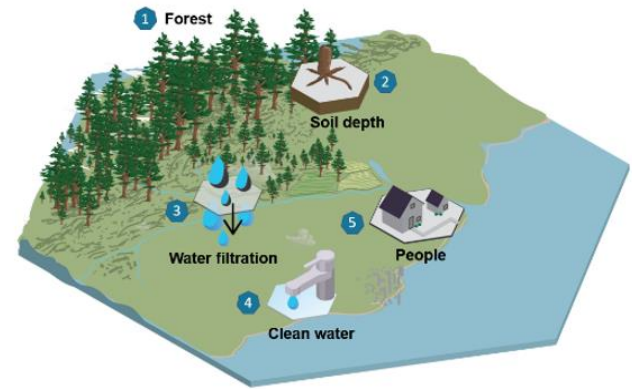

Figure 2: How ecosystem assets generate ecosystem services to beneficiaries in a spatial relationship

**SEEA Ecosystem Accounting**

# WHAT'S NEXT? (4/4)

The underlying assumptions driving this work are:

1. The **level of detail** (disaggregation) of the **structure** plays a crucial role in macroeconomic modeling and simulation.

2. Agent-based modeling can cope efficiently with

   - **Highly disaggregated** structures

   - driven by **heterogeneous** behavioral rules (no equations to be solved, just *if-then* behavioral rules).

There is an awesome amount of <u>static</u> data available. This approach paves the way to ABM <u>dynamic</u> simulation of complex, <u>interacting</u> systems (economics or other) through calibration with static data.

**CONCLUSION**

- The **methodology** presented enables the use of agent-based models for **real-world** economies, as demonstrated by calibrating an ABM model to replicate a SAM, an SUT or an intercountry IOT.

- This Darwinian deployment approach opens the way for the use of **ABM as a tool for policymakers** dealing with complex data from **real macroeconomic** systems.

> In the long term, it provides a **framework** for a possible joint effort to develop **a unified and holistic ABM model for a country, a multiregional or a global system**, including other layers (ecology, climate, social networks...) that can interact with economics.